\documentstyle[twoside,fleqn,espcrc2]{article}
\def\gid{{g_{D^* D \pi}}}
\def\gib{{g_{B^* B \pi}}}
\def\slash#1{\setbox0=\hbox{$#1$}#1\hskip-\wd0\dimen0=5pt\advance
       \dimen0 by-\ht0\advance\dimen0 by\dp0\lower0.5\dimen0\hbox
	 to\wd0{\hss\sl/\/\hss}}
\newcommand{\be}{\begin{equation}}
\newcommand{\ee}{\end{equation}}
\newcommand{\bea}{\begin{eqnarray}}
\newcommand{\eea}{\end{eqnarray}}
\newcommand{\nn}{\nonumber}

\newcommand{\AmS}{{\protect\the\textfont2
  A\kern-.1667em\lower.5ex\hbox{M}\kern-.125emS}}

\title{Couplings of heavy mesons with soft pions in QCD}

\author{P. Colangelo\address{
Istituto Nazionale di Fisica Nucleare, Sezione di Bari, \\
via Amendola n.173, 70126 Bari, Italy}}

\begin{document}

\begin{abstract}
QCD sum rules are used to calculate the couplings of heavy-light quark
pseudoscalar and vector mesons ($D, D^*$ and $B, B^*$) with soft pions, both
for finite and infinitely heavy quark mass. The couplings are also
computed in the framework of a QCD relativistic potential model;
in this approach the
relativistic corrections due to the light quark are relevant.
\end{abstract}

\maketitle

\section{INTRODUCTION}

The strong coupling constant $\gid$, defined by the matrix element
\be
<\pi^+(q)~D^0(q_2) | D^{*+}(q_1,\epsilon )> = \gid
(\epsilon \cdot q)
\label{sda}
\ee
governs the decay $D^{*+} \to D^0 \pi^+$; from the experimental branching ratio
$BR(D^{*+} \to D^0 \pi^+) = 68.1 \pm 1.0 \pm 1.3 \; \%$ \cite {cleo} and from
the measurement $\Gamma(D^{*+}) < 131$ KeV \cite{acc} the upper limit
$\gid \le 20.6$ can be derived.

The analogous coupling constant in the $B$ channel, $\gib$,
describes the contribution
of the  $B^*$ pole to the form factor $F_1^{B \to \pi} (q^2)$
of the semileptonic decay $B \to \pi \ell \nu$:
\be
F_1^{B \to \pi} (q^2_{max}) \simeq {f_{B^*} \over 4}
{g_{B^* B \pi} \over E_\pi + \delta_B } \; ,
\ee
where $f_{B^*}$ is the $B^*$ leptonic decay constant
and $\delta_B=m_{B^*}-m_B$. If the pole of $B^*$
dominates the form factor $F_1^{B \to \pi}$ not only at zero recoil but for all
accessible values of the transferred momentum \cite{pole}, the coupling $\gib$,
together
with $f_{B^*}$, completely determines $F_1^{B \to \pi}$, with interesting
phenomenological consequences for the measurement of $V_{ub}$ using this decay.

In the limit of infinitely heavy quark $b$, $\gib$ scales according to the
relation \cite{nuss}:
\be
\gib= { 2 m_{B^*} \over f_\pi } g \; , \label{ginf}
\ee
where $g$ is a low energy parameter independent of the heavy scale; it acts as
a coupling constant in the effective Lagrangian constructed using heavy quark
and chiral symmetries to describe the interaction of heavy-light quark  $0^-$
and $1^-$ states with the octet of Nambu-Goldstone pseudoscalars \cite{wise}.
The upper limit $g \le 0.7$ can be derived from the experimental bound on
$\gid$; in ref.\cite{def1} the estimate $g \simeq 0.4$ has been given
using heavy-quark chiral symmetry relations and data on the semileptonic
decay $D \to K \ell \nu$.

The couplings $\gid$, $\gib$ and $g$ have been computed in the framework of
nonrelativistic constituent quark models \cite{pot,pot1};
in particular, within this approach the value  $g\simeq 1$
is obtained  \cite{pot1},
which exceeds  the above mentioned experimental bound.

$\gid$ and $g$ are also available from QCD sum rules
\cite{eletski,grozin} and,  recently, from light-cone sum rules \cite{ruckl}.
In ref.\cite{col} we have reconsidered the calculation
in \cite{eletski,grozin}, introducing a number of improvements both in the
theoretical and in the hadronic part of the QCD sum rule, as discussed below.

\section{COUPLINGS FROM QCD SUM RULES}

In order to calculate, for example, the off-shell process
$ B^{*-}(q_1) \to {\bar B^0} (q_2) + \pi^-(q)$
we have considered the time-ordered product of two quark currents
interpolating the $B$
and $B^*$ mesons, between the vacuum and the pion state:
\bea
& & i \int dx <\pi^-(q)| T(V_{\mu}(x) j_5(0) |0> e^{-iq_1x} = \nn \\
& = & A \; q_{\mu} + B \; P_{\mu}
\label{corr}
\eea
where $V_{\mu}={\overline u} \gamma_{\mu} b$, $j_5={\overline b} i\gamma_5 d$
and $P=q_1+q_2$; $A$, $B$ are scalar functions of $q_1^2$, $q_2^2$, $q^2$.
In the soft pion limit ($q \to 0$) and for large Euclidean
momenta $q_1$ and $q_2$ both the invariant functions (we have considered $A$)
can be computed in QCD by the operator product expansion. In our calculation we
have taken into account the contributions proportional
to the quark condensate, the terms proportional to the mixed
quark-gluon condensate (which are missing in \cite{eletski,grozin}) and the
terms proportional to the matrix element
$<\pi (q)|{\overline u} D^2 \gamma_{\mu} \gamma_5 d |0> = -i f_{\pi} m_1^2
q_{\mu}$ (with $m_1^2=0.2 \; GeV^2$ \cite{novikov}).
The following relations have been used:
\bea
<\pi (q)|{\overline u} D_\mu g \sigma G \gamma_5 d |0> &=&
{m_0^2 <\bar q q> \over f_\pi} q_\mu \nn \\
<\pi (q)|{\overline u} D^\alpha g G_{\alpha \mu} \gamma_5 d |0> &=&
- {i \over 4} {m_0^2 <\bar q q> \over f_\pi} q_\mu \; .\nn\\
\eea
On the other hand, a dispersive representation can be given for $A$:
\be
A(0,q_1^2,q_2^2)={1\over{\pi^2}} \int ds \; ds' {{\rho(s,s')}\over{(s-q_1^2)
(s'-q_2^2)}} \;
\label{intg}
\ee
with the spectral function $\rho$ expressed in terms of hadronic states.
In the region
$ m_b^2 \le s, s^\prime \le s_0$ (with $s_0$ a small threshold)
$\rho(s,s^\prime)$
gets contribution from the $B$ and $B^*$ poles, only.
Above the threshold other  contributions must be taken into account
(in addition to the hadronic continuum)
 which are not suppressed by the borelization procedure
since, in the soft pion limit, a single Borel transform
in the variable $q_1^2=q_2^2$ must be performed; these resonance-continuum
contributions are called "parasitic terms".
By studying models for such terms, we have observed that they generally
contribute to the Borel transformed sum rule for $\gib$ as follows:
\be
{\gamma \over {M^2}} + d_0 + d_1 e^{- {\delta/M^2} } =
 e^{\Omega/M^2} \; f^{QCD} (M^2)
\label{sr}
\ee
where $M$ is the Borel parameter,
$\delta=s_0 -m_B^2$,
$\gamma=f_B~f_{B^*}~m_B^2~\gib (3 m^2_{B^*} + m^2_B)/4~m_b~m_{B^*}$ and
$\Omega = (m_B^2+m_{B^*}^2-2m_b^2)/2$; $f^{QCD}$ is the Borel transformed
QCD side of the sum rule (the expression can be found in \cite{col}).

The terms
$d_0$ and $d_1$ parametrize the parasitic contributions:
they do not
depend on $M$ and therefore can be removed performing derivatives in
$M^2$.

The numerical analysis has been carried out using the parameters:
$m_c=1.35 \; GeV$, $m_b=4.6 \; GeV$, the standard values of the condensates
and the effective thresholds
$s_0=6-8 \; GeV^2$ in the case of the $c$ quark and
$s_0=32-36 \; GeV^2$ in the case of the $b$ quark.
The result is:
\bea
f_B \; f_{B^*} \; \gib & = &  0.56 \pm 0.12 \; GeV^2  \label{res1} \nn \\
f_D \; f_{D^*} \; \gid & = &  0.34 \pm 0.08 \; GeV^2   \label{res2} \nn \\
{\hat F}^2 \; g & = & 0.035 \pm 0.008 \; GeV^3 \; ; \label{res}
\eea
in the last equation the result concerning the limit
$m_b \to \infty$ is reported, with ${\hat F}$ given by
${\hat F}= f_B \sqrt{m_B}$ in this limit.

The results in (\ref{res}) agree with the light-cone sum rules calculation
in \cite{ruckl}.

The prediction for $\gid$, $\gib$, and $g$ can be given once the values of the
leptonic constants $f_D$ etc. are put in (\ref{res}).
Using the leptonic constants obtained by two-point function QCD sum rules
without radiative corrections (notice that the calculation for the strong
couplings has been carried out at the order ${\cal O} (\alpha_s)=0$)
we get: $\gid = 9 \pm 2$, $\gib = 20 \pm 4$ and $g = 0.39 \pm 0.16$.
On the other hand,
if radiative corrections at the order ${\cal O} (\alpha_s)$
are included in the calculation of the leptonic constants we obtain:
$\gid = 7 \pm 2$, $\gib = 15 \pm 4$, $g = 0.21 \pm 0.06$.  Within the
uncertainties, there is agreement for $\gid$ and $\gib$; the difference between
the values for $g$ reflects the important role of radiative corrections in the
determination of $f_B$ in the limit $m_b \to \infty$.

Using the central values for $\gid$ and $\gib$  we predict:
$\Gamma(D^{*+})=10-17 \; KeV$ and
$F_1^{B \to \pi} (0) = 0.30-0.40$ (in the hypothesis of dominance of the $B^*$
pole for this form factor).

\section{COUPLINGS FROM A QCD RELATIVISTIC POTENTIAL MODEL}

The value $g \simeq 0.2 - 0.4$ obtained by QCD sum rules must be compared with
the outcome of constituent quark models: $g \simeq 1$ \cite{pot1}.
Since in these models
the light quark is treated as a nonrelativistic particle,
one could wonder about the role of relativistic corrections.

Relativistic
effects can be taken into account in the framework of the potential model
described in \cite{col1}; in this model the  quark kinematics is
relativistic, and the interquark interaction is described by the Richardson
potential, linear at large distances and coulombic (with QCD corrections)
at short distances.

Within this approach the following equation has been
derived for $g$ \cite{col2} (in the limit $m_b \to \infty$):
\bea
g  &=&  {1 \over 4 m_B} \int_0^\infty dk |\tilde{u}_B(k)|^2 \times \nn \\
&\times& {E_q + m_q \over E_q} \left[ 1-{k^2 \over 3 (E_q+m_q)^2} \right] \;,
\label{ginf1}
\eea
where $\vec k$ is the relative quark-antiquark momentum inside the meson,
$\tilde{u}_B(k)$ is the $B$ and $B^*$ wave function (we neglect the spin
splitting between $B$ and $B^*$ which is of the order $1/m_b$),
$E_q=\sqrt{k^2 + m_q^2}$ ($q$ is the light quark); the wave function
normalization is:
$ \int_0^\infty dk |\tilde{u}_B(k)|^2=2 \; m_B $.

In the non relativistic limit
$E_q \simeq m_q \gg k$ one obtains:
$g=1$.
This is the result of constituent quark models; notice that,
in this limit, eq.(\ref{ginf1}) is similar to the expression for $g$ derived by
Kamal and Xu in \cite{pot1} in the framework of the Bauer-Stech-Wirbel model.

On the other hand, if one considers the limit $m_q \to 0$
(our fit of the meson masses fixes the light quark mass to
the value $m_q=38 \; MeV$)
the equation for $g$ gives the result:
$g={1 \over 3}$, in agreement with QCD sum rules.
This allows us to conclude that the reduction of the value of $g$
from the result of non relativistic constituent quark models
finds an explanation (as observed also in \cite{od})
in the effects of the relativistic motion of the light quark.

\vskip 15pt
{\bf ACKNOWLEDGMENTS \\}
I thank A.Deandrea, F.De Fazio, N.Di Bartolomeo, F.Feruglio, R.Gatto and G.
Nardulli for their collaboration in the studies described in this talk.
Thanks are also due to R.Casalbuoni and N.Paver for interesting discussions.

\end{document}